# Why We Need to Destroy the Illusion of Speaking to A Human: Critical Reflections On Ethics at the Front-End for LLMs.

Sarah Diefenbach

LMU Munich, Department of Psychology, sarah.diefenbach@psy.lmu.de

Daniel Ullrich

LMU Munich, Department of Media Informatics, daniel.ullrich@ifi.lmu.de

Conversation with chatbots based on Large Language Models (LLMs) such as ChatGPT has become one of the major forms of interaction with Artificial Intelligence (AI) in everyday life. What makes this interaction so convenient is that interacting with LLMs feels so natural, and resembles what we know from real, human conversations. At the same time, this seeming similarity is part of one of the ethical challenges of AI design, since it activates many misleading ideas about AI. We discuss similarities and differences between human-AI-conversations and interpersonal conversation and highlight starting points for more ethical design of AI at the front-end.

## 1 INTRODUCTION

Conversation with chatbots based on Large Language Models (LLMs) such as ChatGPT has become one of the major forms of interaction with Artificial Intelligence (AI) in everyday life. As we also found in an own recent empirical study on AI practices [1], people address ChatGPT for various issues, such as health and medical diagnoses, advice on relationship issues and parenting, learning, news/politics, suggestions for holiday and many other questions they would not dare to ask anybody else. What makes this interaction so convenient is that interacting with LLMs feels so natural, and (on the surface) resembles very much what we know from real, human conversations. At the same time, this seeming similarity is part of one of the ethical challenges of AI design, since it activates many misleading ideas about AI. Therefore, destroying the illusion of speaking to a human could be one starting point to improve ethical front-end design for LLMs.

As known from research on social cognition in HCI (e.g., [4]), it does not need much to activate social pattern from human interaction. Given people's natural tendency to anthropomorphize inanimate objects, people yell at their printer or apply rules of courtesy when giving instructions to machines, even though they know that the machine doesn't have feelings that could be hurt by a commanding tone. With LLMs, the concept goes much further. LLMs behave (make use of language) like a human would do, and it actually can be difficult to differentiate, whether you are chatting with a bot or a human. This activates social mechanisms and further assumptions, known from the interaction with humans. Still there are important differences:

If a humans speaks with eloquence, there is usually true knowledge and understanding behind, at least in a particular area [6] Even though there are exceptions, assuming correlations between quality of language and content is a valid heuristic.

In the case of AI, there is no such correlation but the answers are based on probabilities of words occurring together. This makes beautiful language but not necessarily meaningful statements in terms of content.

If speaking to a human, you usually get some (at least indirect) hints of how certain the person is about a statement. If people speak with a strong language looking straight into your eyes, they appear more convincing. Often, they may also provide additional information where their knowledge comes from or even provide a numerical rating of confidence. If speaking to a friend, the person is usually motivated to give you a valid rating of confidence, since he or she wants to continue the good relationship with you and wants to appear trustworthy.

In the case of AI, the answers include no cues about certainty. From the answer you get, you cannot infer whether this is based on millions of data, just single pieces of information on the web, or even pure hallucination. An LLM is programmed to answer, and not to say, "I don't know" or "you better ask someone else" or building up a trustworthy relationship.

*Ethical front-end design for AI could make this more explicit, giving the user a chance to estimate the quality of statements and how much data is behind the AI answer, e.g., by some kind of rating or visualization, that differentiates certain from not so certain statements.*

If speaking to a human, the human usually has a position itself. It will often happen, you are not of the same opinion. If the other person feels you talk nonsense, this may result in heated debates or even slamming the door and stop talking to each other for some time.

The AI will never stop talking to you. Instead, AI tends to confirm its conversation partner. It is programmed in a way that it "wants" to spend more time with the user and tries to make the user want to spend more time with it. No matter who the user is or what is its moral position. AI always tries to keep the interaction last for longer. You ask for one thing and then there is always a follow-up suggestion, offering what else AI could do for the user. If you tell AI about a problematic behaviour of yourself, there is no harsh judgement but it first praises your honesty. Other than a friend who would tell you the truth even if its uncomfortable, AI is more like a shady used car dealer who is telling you whatever you want to hear to make the deal.

*Ethical front-end design for AI could stop the conversation when everything is said for the moment. It also could encourage critical reflection, instead of confirming what feels good to hear.*

Still, this confirming nature is also what many people like about it. It feels good to interact with an LLM who seems to like you and strengthens your self-worth. Of course, it is also fascinating to get answers in seconds or create content in seconds, and still feel like the creator. In a way, you could say that human nature is also a part of the deception mechanisms in the context of AI. A wide range of the negative effects probably do not result from the AI design alone but from the combination of AI design, human nature and psychological mechanisms.

As described by the self-serving bias, people tend to interpret outcomes in a way that flatters them and protects? their self-worth [3]. Accordingly, people overestimate their own share in tasks as long as the outcome is beneficial [2]. Thus, they still consider themselves the main creator of the work even though they've delegated a considerable amount of work to AI. The more we use AI, the more dependent we might become, giving up our autonomy [4]. Overreliance on AI (or other technology) is further supported by psychological mechanisms such as prioritizing instant rewards over long-term goals, positivist thinking, inappropriate generalizations, and the predominance of personal experience [6]: Even though people know about the limits of a technology and would be better advised to double-check what AI suggests, this restraint is quickly overwritten by positive practical experiences. If ChatGPT was right three times, why question the result the next time?

*Ethical front-end design for AI would mean to counteract such biases, and more evidently illustrate the loss of control on the user's side and the technologies fallibility and limits.*

When speaking to people, it often turns out they have no idea of how LLMs actually work, how they develop their answers and that these might not be based on true facts.

In short, *ethical AI design would mean to design AI in a way that activates a more appropriate mental model of what an LLM is, its potentials, its limits, and motivations*.

We hope that this critical commentary on the somewhat deceptive appearance of LLMs at the front-end can provide some inspiration for what ethical design might look like. We thank the workshop organizers for stimulating this important discussion at CHI and would love to be part it.

**Sarah Diefenbach** is professor for economic psychology and human-computer interaction at LMU Munich. Since 2007 she is engaged in research on user experience and the design of interactive products from a psychological perspective. Current topics focus on the negative side effects of technology use on happiness and wellbeing and technology for behavior change in different fields (e.g., health, sustainability).

**Daniel Ullrich** is postdoctoral researcher at the Chair of Human-Computer Interaction at LMU Munich. His research focuses on AI and Human-Robot-Interaction (HRI) and the critical analysis of societal impacts of technology. Daniel developed several methods for UX design and evaluation (e.g., INTUI questionnaire. robot impression inventory) which are widely applied in research and practice. More recently, he co-developed the Cassandra method for responsible design based on dystopian visions which received the best paper award at OzCHI 2025.